\documentclass{an}
\usepackage{graphicx}
\usepackage{times}
\begin{document}
\Pagespan{999}{}
\Yearpublication{2015}%
\Yearsubmission{2015}%
\Month{1}%
\Volume{999}%
\Issue{999}%

\title{The Age and Interior Rotation of Stars from Asteroseismology}

\author{Conny Aerts\thanks{Corresponding author:
  \email{Conny.Aerts@ster.kuleuven.be}}}
\titlerunning{The Age and Internal Rotation of Stars from Asteroseismology}
\authorrunning{Conny Aerts}
\institute{
Institute of Astronomy, Department of Physics \& Astronomy, 
KU\,Leuven, B-3001 Leuven, Belgium
\and 
Department of Astrophysics, IMAPP, Radboud University Nijmegen, the Netherlands
\and
Kavli Institute for Theoretical Physics, University of California, Santa
Barbara, CA 93106, USA
}

\received{?? February 2015}
\accepted{?? February 2015}
\publonline{later}


\abstract{%
We provide a status report on the determination of stellar ages from
  asteroseismology for stars of various masses and evolutionary stages. The
  ability to deduce the ages of stars with a relative precision of typically 10
  to 20\% is a unique opportunity for stellar evolution and also of great value
  for both galactic and exoplanet studies. Further, a major uncalibrated
  ingredient that makes stellar evolution models uncertain, is the stellar
  interior rotation frequency $\Omega(r)$ and its evolution during stellar life.
  We summarize the recent achievements in the derivation of $\Omega(r)$ for
  different types stars, offering stringent observational constraints on
  theoretical models. Core-to-envelope rotation rates during the red giant stage
  are far lower than theoretical predictions, pointing towards the need to
  include new physical ingredients that allow strong and efficient coupling
  between the core and the envelope in the models of low-mass stars in the
  evolutionary phase prior to the core helium burning. Stars are subject to
  efficient mixing phenomena, even at low rotation rates. Young massive stars
  with seismically determined interior rotation frequency reveal low
  core-to-envelope rotation values.}
 
\maketitle

\section{Asteroseismology: the New Route for Stellar Physics}

Contemporary stellar structure and evolution theory still has
several open questions, the answers having vast implications for exoplanetary,
supernova, and (extra)galactic science.  One major uncalibrated quantity is the
rotation frequency in the stellar interior throughout stellar life. Also
the level of chemical mixing inside stars is hard to judge from surface
abundance measurements. While models of stars not too different from the
Sun in terms of mass, rotation, and evolutionary stage are well scalable from the
solar model, this is far less so for stars with appreciably different properties such as
high mass, rapid rotation, strong wind, evolved status, etc.

\begin{figure}[t!]
\begin{center}
\rotatebox{270}{\resizebox{8.cm}{!}{\includegraphics{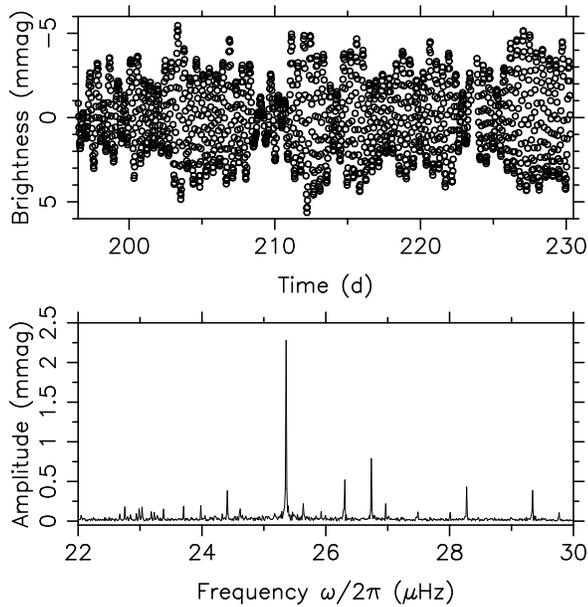}}}
\end{center}
\caption[]{Top: excerpt from the {\it Kepler\/} light curve of the 
g-mode pulsator KIC\,9210943; bottom: frequency spectrum in the
  range of the dominant modes based on the full light curve spanning 1470\,d.}
\label{LC}
\end{figure}

A recent driver to improve stellar physics
is the availability of long-duration high-cadence
quasi-uninterrupted white-light space photometry with a precision of $\mu\,$mag
assembled for thousands of stars by the European CoRoT (operational from 2006 to
2012, Auvergne et al.\ 2009) and the NASA {\it Kepler\/} (operational from 2009
to 2013, Koch et al.\ 2010) satellites. This offered the opportunity to confront
stellar models with {\it asteroseismic\/} data.  This is done by computing the
predicted spectrum of normal oscillation modes from theoretical models, by
considering small perturbations to the equations of stellar structure, usually
in the approximation of a spherically symmetric star.  Normal modes are either
{\it pressure\/} (p-)modes or {\it gravity\/} (g-)modes, depending on whether
the pressure force, respectively buoyancy, is the dominant restoring force.
Pressure modes probe the stellar envelope while g-modes tune the inner part of
star. Such seismic diagnostics are far more suitable to probe stellar
interiors compared to measurements of surface quantities.  Figure\,\ref{LC}
shows part of a typical {\it Kepler\/} light curve for a newly discovered g-mode
pulsator, as well as the frequency spectrum based on the entire light curve in
the range of maximum amplitude.  An extensive
description of the method of asteroseismology is available in Aerts et al.\
(2010).

Seismic modelling consist of searching for a theoretical stellar model whose
frequency spectrum matches the observed one in the bottom panel of
Fig.\,\ref{LC} to within the measurement errors.  A basic scheme of how forward
seismic modelling works is available in Aerts (2013, Section1 and Fig.\,1) and
is hence not repeated here. For each of the detected pulsation frequencies,
$\omega_{n\ell m}$, identification of the radial order $n$ and spherical
wavenumbers, $(\ell,m)$ representing the nodes of the eigenmode, is necessary
in order to perform meaningful forward modelling.  Once mode
identification has been achieved, the seismic modelling is usually based on the
frequency values of zonal $(m=0)$ modes, which represent standing waves that are
not affected by the stellar rotation in the case of a spherically symmetric
star.  The modes with $m\neq 0$ represent waves travelling in the direction of
rotation (prograde modes) or opposite to it (retrograde modes) and can in
principle be used to derive the internal rotation frequency $\Omega (r)$ after a
good seismic model based on the zonal modes has been achieved. We come back to
this opportunity in Section\,\ref{SectionRotation}.

In the limits of high- or low-frequencies, identification of $(n,\ell,m)$ can
often be achieved from frequency patterns. The {\it large frequency
  separation\/} of p-modes of high radial order $n$, e.g., is the difference in
frequency of modes with the same degree but of consecutive radial order. This
separation equals the inverse of twice the sound travel time between the stellar
surface and centre and offers a high-precision measurement of the mean density
of the star. {\it Stochastically excited modes\/} are triggered by the
convective envelope of stars and these modes appear as Lorentzian peaks in
Fourier space. They fulfill so-called {\it scaling relations\/} based on the
large frequency separation and the {\it frequency of maximum power}, allowing to
estimate the stellar mass and radius when the effective temperature is known
(Kjeldsen \& Bedding 1995). The validity of these relations was already
confirmed for solar-type stars with pulsations detected in ground-based
radial-velocity data (e.g., Bouchy \& Carrier 2001; Bedding et al.\ 2004;
Kjeldsen et al.\ 2005) and have now seen widespread application to extensive
ensembles of unevolved solar-type stars (e.g., Chaplin et al.\ 2014) and red
giants (e.g., Hekker et al.\ 2011; Huber et al.\ 2011; Kallinger et al.\ 2012;
Stello et al.\ 2013).

For g-modes, a natural diagnostic to consider in forward modelling is the {\it period
spacing\/} between high-order low-degree modes of consecutive radial order. This
period spacing is connected with the detailed properties of the buoyancy
frequency inside the star and probes the deep interior.  As a star grows old, it
builds up chemical composition gradients due to its nuclear burning
stages. Hence, {\it trapping\/} of g-modes occurs in the deep interior and this
gives rise to deviations from uniform period spacings (e.g., Brassard et al.\ 1992;
Miglio et al.\ 2008).

A major discovery based on {\it Kepler\/} light curves is the occurrence of
dipole mixed modes in red giants (Beck et al.\ 2011; Bedding et al.\
2011). These were predicted theoretically by Dupret et al.\ (2009) and have a
p-mode character in the star's envelope while g-mode properties in the core
region. They allow the combined use of the large frequency separation and the
period spacing as an important diagnostic to deduce if the red giant is in the
hydrogen shell burning phase or rather in the core helium burning phase (Bedding
et al.\ 2011). Such a distinction cannot be made without seismic data
because the surface properties of red giants are alike.

Below, two important ``deliverables'' of asteroseismology for astrophysics are
emphasized. It concerns the age and interior rotation profile of
stars. Asteroseismology can offer these quantities with a precision that no
other method can.  By choosing this focus, this paper by no means gives an
overview of all the results that have been achieved in asteroseismology lately.
Moreover, the text is heavily biased towards modelling results. These rely on
the outcome of asteroseismic data analyses.  Hence our text does not bring
justice to the hundreds of instrumental and observational studies upon which the
seismic inferences are based. Extensive earlier reviews of asteroseismology in
the space era are available in, e.g., Michel et al.\ (2008), Gilliland et al.\
(2010), Christensen-Dalsgaard \& Thompson (2011), and Chaplin \& Miglio (2011).

\section{Stellar Ages}
\label{SectionAges}

Stellar ages are very hard to determine with high precision, while this quantity
is of vast importance in astrophysics.  The formation and evolution of exoplanet
host stars, e.g., directs the formation and evolution of their planets. On the
other hand, galactic archeology and nearby cosmology require to identify
the oldest well-accessible stellar populations in various regions of the Milky
Way and in neighbouring galaxies. This can now be tackled seismically from
red giant studies.

While the recent applications of {\it gyrochronology\/} of rotationally modulated active
stars is highly successful (e.g., Epstein \& Pinsonneault 2014; Meibom et al.\
2015), this method of aging is limited to stars in a very narrow low-mass range,
typically between 0.8 and 1.4\,M$_\odot$. Asteroseismic aging can be done for
all mass ranges, whenever several non-radial pulsation mode frequencies can be
detected and identified.  Garci\'a et al.\ (2015) bridged gyrochronology and
asteroseismic aging for lower-mass stars and found compatible results, thus
calibrating the gyrochronology relation seismically. However, they stressed that
the age-rotation-activity relations seem quite different for the hotter dwarfs
and subgiants.  For this reason, and because it works for all masses, seismic
aging is a major asset.

\subsection{Aging of Low-Mass Solar-like Pulsators}

\subsubsection{Core Hydrogen Burning Stage}
In contrast to the mass and radius, estimation of the stellar age cannot be
achieved model-independently from the scaling relations based on the large
frequency separation and the frequency of maximum power of solar-like pulsators.
It requires the additional measurement of the so-called {\it small frequency
  separation}, which is the frequency difference between adjacent radial and
quadrupole zonal modes differing by 1 in radial order. This
seismic quantity is connected with the sound speed gradient inside the star and
hence tunes the regions where nuclear burning has taken place in the past and/or
is occurring presently. Coupled to the evolutionary track for the seismic mass
deduced from the scaling relations, the small frequency separation delivers the
stellar age, {\it for a particular metallicity and for the chosen input physics
  of the stellar models}.  This method was applied to a large sample of
more than 500 solar-like core hydrogen burning stars and provided relative age
precisions of 10 to 15\% (Chaplin et al.\ 2014).

Detailed stellar modelling of case studies taking into account the frequencies
themselves rather than just a few average quantities, offers a much
better way to derive the age of low-mass stars. This was demontrated by Lebreton
\& Goupil (2014), who presented an extensive study of the exoplanet host star
HD\,52265 based on four months of CoRoT data.  In their study, they quantified
the impact of various assumptions for the input physics and free parameters
(such as the mixing length), and their uncertainties, used in the stellar
models on the derived age. 
This allowed them to obtain relative precisions on the age, mass, and
radius that are smaller than those obtained from the large and small frequency
separations alone, although the frequency precision for their case study was
limited.  Metcalfe et al.\ (2014) performed a similar but far more extensive study of 42
solar-like stars based on nine months of {\it Kepler\/} data. They found that
forward modelling based on the individual frequencies typically doubles the
precision of the asteroseismic radius, mass, and age compared to estimates based
on the scaling relations or on the large and small frequency separations. The
full four-year {\it Kepler\/} light curves have yet to be explored in terms of detailed 
forward modelling based on the individual frequencies.

Lebreton \& Goupil (2014) stressed the tight correlation between the initial
helium abundance and mass of the star. This was taken up by Verma et al.\ (2014)
in the case of the solar analogs 16\,Cyg A and B, monitored during 2.5\,yr by
the {\it Kepler\/} mission. These data allowed the authors to determine the
current helium abundances of these stars to lie between 0.231 and 0.251 for
16\,Cyg\,A and between 0.218 and 0.266 for 16\,Cyg\,B.  This resulted from
detailed modelling of the pulsation frequencies, keeping in mind that the helium
ionization zone leaves particular signatures on the oscillation frequencies that
can be measured from frequency values of sufficiently high precision, as offered
by the 2.5\,yr time base of the {\it Kepler\/} data. These signatures are
connected with so-called {\it acoustic glitches}. These are due to regions in
the stellar interior where the sound speed experiences an abrupt variation due to a
local change in the stratification. Acoustic glitches occur at the boundaries
between radiative and convective regions, as well as in the partial ionisation
layers, especially those of hydrogen and helium. Such a glitch introduces an
oscillatory component in the pulsation frequencies of the star.

Mazumdar et al.\ (2014) exploited the measured acoustic glitches of 19
solar-type stars based on nine months of {\it Kepler\/} data, along with the large
and small frequency separations to locate the base of the convective envelope
and the position of the second helium ionisation zone. They found good agreement
with state-of-the-art models for these 19 stars. Future applications of actual
foward modelling based on frequency fitting may offer the potential to constrain
the metallicity and chemical mixture of solar-like stars, by
exploiting the full {\it Kepler\/} data sets.

\subsubsection{Evolved Stages}

While the small frequency separation is an appropriate age indicator for the
core hydrogen burning phase, this is no longer the case for the sub-giant phase.
Indeed, subgiants have a rather inert helium core that hardly undergoes
evolutionary changes. On the other hand, the stellar core starts shrinking when
the star climbs up the red giant branch. The small frequency separation can
then again be used to tune the age in that stage, given that it is sensitive
to the consequences of the hydrogen shell burning (e.g., White et al.\,2011).

During the red giant branch evolution, the small frequency separation changes
gradually until helium ignition through the helium flash occurs. After the
flash, when core helium burning occurs in equilibrium after several sub-flashes,
the small frequency separation is typically some 10\% larger than the one on the
red giant branch (e.g., Corsaro et al.\ 2012; Kallinger et al.\ 2012).  The
theory of stellar evolution during core helium burning on the horizontal branch
is not yet well enough understood to interpret changes in the small frequency
separation in terms of the stellar age alone. Indeed, various physical phenomena
such as core overshooting, atomic diffusion, turbulent and/or diffusive 
mixing, mass loss, and
core rotation are active together, while they are yet poorly understood and
remain uncalibrated. Each of them has an effect on the small frequency
separation. For this reason, Kallinger et al.\ (2012) suggested to use the
detailed properties of the three dominant radial modes in red giant frequency
spectra, including their phase behaviour, as an appropriate age proxy that can
also be applied to ground-based spectroscopic seismic data of bright pulsating
red giants. Nevertheless, as stressed by Mosser et al.\ (2014) and
following Bedding et al.\ (2011), a measurement of the period spacing of dipole
mixed modes along with the large frequency separation of p-modes offers the best
way to derive the evolutionary stage of low-mass stars, all the way from the
sub-giant phase to the end of the core helium burning.

Another type of core helium burning stars for which forward seismic modelling
has been achieved are the subdwarf B (sdB) stars. These objects are situated at
the blue end of the horizontal branch and are responsible for the observed
UV-upturn of early-type galaxies. It remains unclear how the sdB stars have lost
their hydrogen envelope during the red giant branch phase, but binarity is held
responsible for it, given that half of the sdB stars reside in close binaries
with a white dwarf or low-mass core hydrogen burning companion (Heber 2009).
While red giant pulsation are excited by the convection in the outer envelope,
sdB stars have no such envelope but may experience pulsation modes excited in
the partial ionisation layers of iron-like elements.  Seismic modelling of sdB
stars is a major asset to tune the physics of core helium burning stars, but was
so far limited to single-star models starting from the onset of core helium
burning, irrespective of the previous evolutionary history. Age estimates are
then found with respect to the zero-age horizontal branch (ZAHB). The frequency
spectra of sdB pulsators do not always permit to identify the mode numbers
$(n,\ell,m)$. In that case, these are estimated along with the modelling, by
considering all modes of low degree to fit the frequencies. While this may seem
arbitrary, this procedure was demonstrated to be fully appropriate from the
binary versus asteroseismic modelling of the pulsating reflection eclipsing
binary sdB pulsator PG\,1336-018 (Vu\v{c}kovi\'c et al.\ 2007; Van Grootel et
al.\ 2013).

Van Grootel et al.\ (2010a) performed the first detailed forward modelling of a
g-mode sdB pulsator, KPD\,0629-0016, observed with the CoRoT mission during
21\,d.  They found a total mass of $M_{\rm tot}=0.471\pm 0.002\,$M$_\odot$, a
hydrogen envelope mass of $\log\,(M_{\rm env}/M_{\rm tot})=-2.42\pm 0.07$, and
an age of 42.6$\pm$1.0\,Myr since the ZAHB.  These results differ
from those for the younger sdB pulsator KPD\,1943+4058 observed with the {\it
  Kepler\/} mission in terms of total mass, but are similar for the hydrogen
envelope: $M_{\rm tot}=0.496\pm 0.002\,$M$_\odot$, $\log\,(M_{\rm env}/M_{\rm
  tot})=-2.55\pm 0.07$, for an age of 18.4$\pm$1.0\,Myr since the ZAHB (Van
Grootel et al.\ 2010b).  Charpinet et al.\ (2011) modelled a third sdB pulsator,
KIC\,2697388, but could not pinpoint a unique value for the total and envelope
masses, because of a degeneracy in two families of solutions of equal quality in
terms of seismic modelling.  Nevertheless, they derived an upper age limit of
$\sim\,55\,$Myr since the ZAHB and came to the conclusion that 
the {\it seismically modelled sdB pulsators hint
  towards the need of extra mixing since the ZAHB}, caused by e.g.\ core
overshooting and/or differential rotation, compared to the models based on the
Schwarzschild criterion of convection to define the inner convective core.  As
we discuss below, this conclusion also holds for massive stars with a
convective core during the core hydrogen burning.

\begin{figure*}
\begin{center}
\rotatebox{0}{\resizebox{8.53cm}{!}{\includegraphics{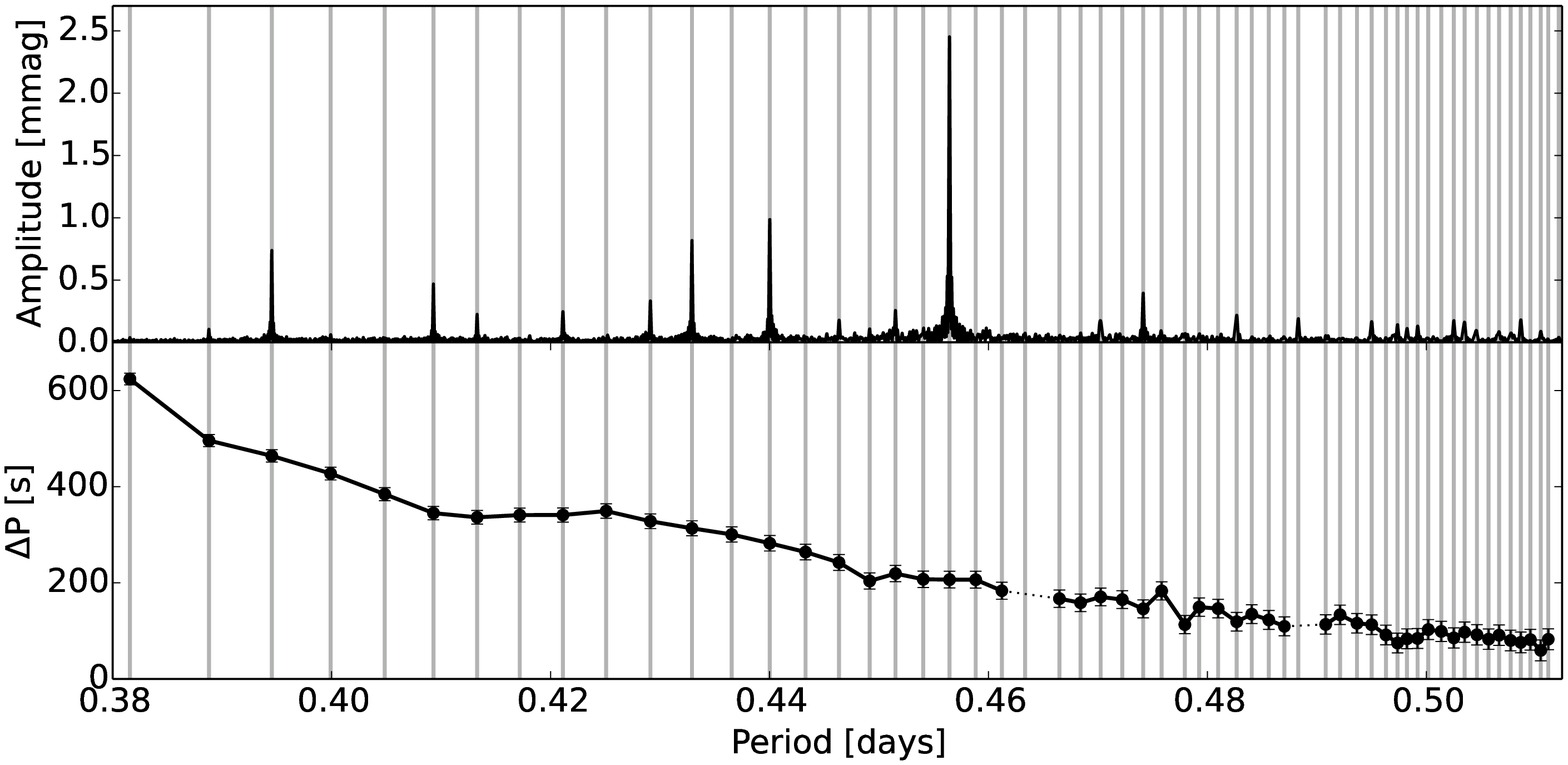}}}
\rotatebox{0}{\resizebox{8.53cm}{!}{\includegraphics{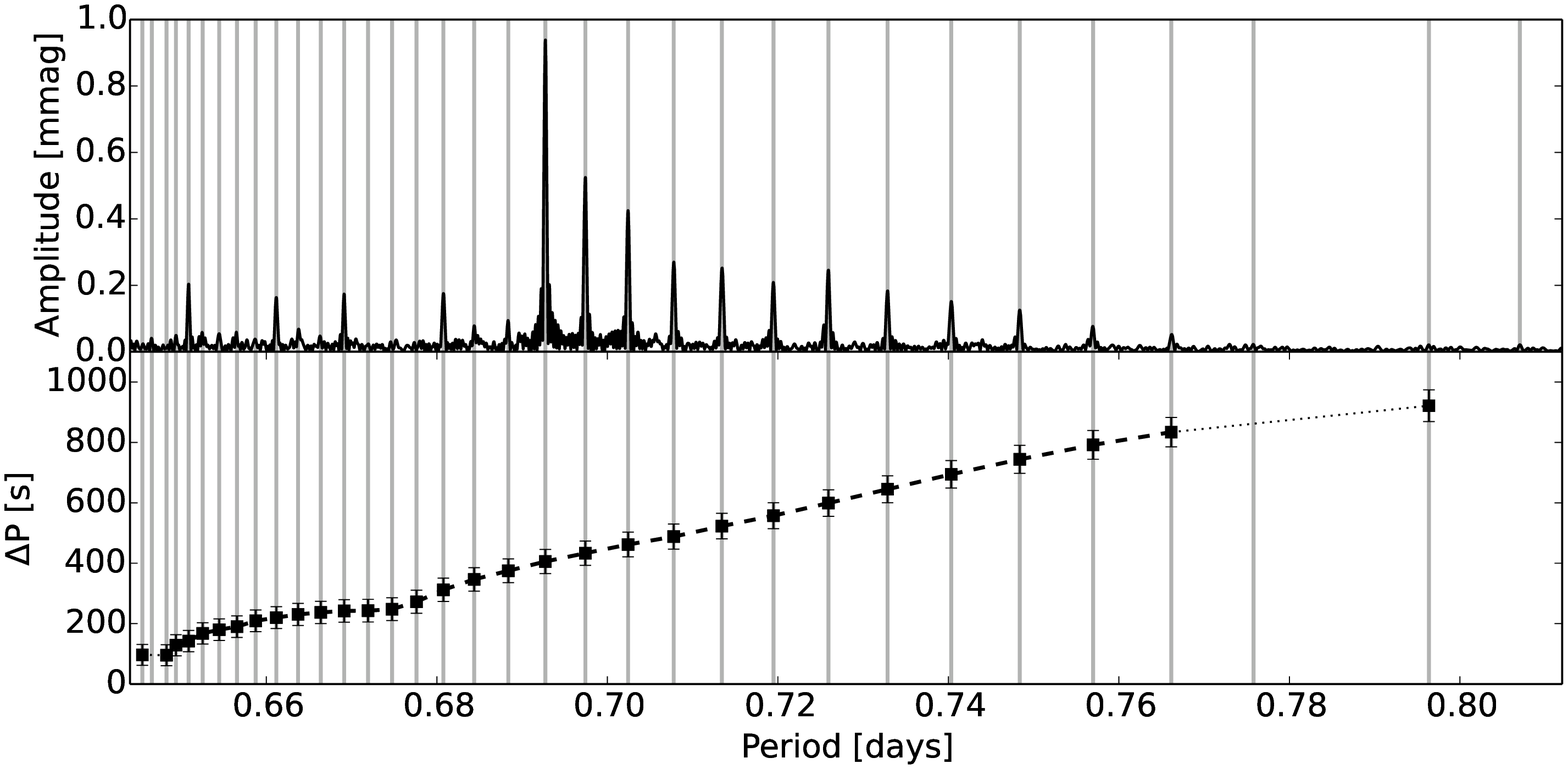}}}
\end{center}
\caption[]{Period spacings of prograde (left) and retrograde (right) dipole
  modes in a {\it Kepler\/} F-type pulsator from the sample by Tkachenko et al.\
  (2013). Significant frequencies belonging to the same spacing pattern are
  indicated with vertical grey lines.  Figure reproduced from Van Reeth et al.\ (2015b).}
\label{Timothy}
\end{figure*}

\subsection{Aging of Massive Pulsators}

Not only sdB stars, but numerous classes of stars, in various evolutionary
phases, experience so-called self-driven non-radial pulsation modes excited by a
heat mechanism, usually due to an opacity bump occurring in partial ionisation
layers --- see, e.g., Aerts et al.\ (2010), Chapter 2, for an overview of all
those classes and their pulsational properties. The highest-mass star for which
such heat-driven non-radial pulsations have been modelled seismically concerns
the slowly rotating O9V star HD\,46202, leading to mass of
24$\pm$0.8\,M$_\odot$, a core overshooting value of 0.10$\pm$0.5 local pressure
scale heights, and an age of 4.3$\pm$0.5\,Myr (Briquet et al.\ 2011). Unlike
stochastically-excited solar-like pulsations, the heat-driven modes are not
damped and so their frequencies give rise to delta peaks in Fourier space. This
makes it easier to determine them with high precision compared to the Lorentzian
frequency peaks caused by the continously damped and re-excited stochastic
solar-like pulsations. On the other hand, heat-driven modes do not necessarily
occur in a frequency regime that gives rise to frequency separations or period
spacings and lack of unambiguous identification of $(n,\ell,m)$ corresponding
with the detected mode frequencies may impose a serious limitation, preventing
seismic modelling.

A new way forward in seismic modelling and aging for intermediate- and high-mass
stars in core hydrogen burning, based on period spacings due to high-order
gravity modes, was initiated by Degroote et al.\ (2010). They did not only
detect the period spacing of 9,450\,s due to dipole modes in the B3V Slowly
Pulsating B (SPB) star HD\,50230 of $\sim$7\,M$_\odot$ observed with CoRoT, but
they also derived periodic deviations from the constant spacing with a
periodicity of 2,450\,s and an amplitude of some 240\,s, decreasing with
increasing mode period. This is clearly the signature of {\it a smooth gradient
  of chemical composition due to diffusive mixing\/} inside the star, because
instantaneous near-core mixing would leave much sharper features in the period
spacing (Miglio et al.\ 2008), given that forward modelling led to the conclusion
that the star has already consumed about 60\% of its initial hydrogen.  An
absolute age could not be derived for this star, due to too poor constraints on
the core overshoot value for which only a lower limit of 0.2 local pressure
scale heights could be concluded.

Detections of period spacings have recently also been announced for F-type
$\gamma\,$Dor pulsators, both in the case of slow rotators with average spacings
of a few thousand seconds (Bedding et al.\ 2014) and for modest to fast rotators
whose period spacings are affected strongly by rotation and have values of
only a few hundred seconds (Van Reeth et al.\ 2015a,b --- see
Fig.\,\ref{Timothy} showing one of the sample stars).  Such decreased values for
the spacings are a clear signature of the effect of rotation on the frequency
patterns, as theoretically expected (Bouabid et al.\ 2013). Moreover, the smooth
downward and upward patterns without deep dips as observed in
Fig.\,\ref{Timothy} are a signature of diffusive mixing inside the star,
inhibiting strong mode trapping, as outlined by Bouabid et al. (2013). While the
sample of $\gamma\,$Dor stars by Tkachenko et al.\ (2013) is still subject of
forward seismic modelling based on the observational findings in Van Reeth et
al.\ (2015b), it is already clear that the detection of prograde and/or
retrograde dipole period spacings as it has been achieved for 45 $\gamma\,$Dor
stars of the sample, one of which shown in Fig.\,\ref{Timothy}, will offer an
excellent opportunity to model the interior rotation, chemical mixing, and aging
of stars in the mass range of 1.5 to 2.5\,M$_\odot$ in the near future.

To conclude this section, we stress that mixing inside stars, particularly when
it occurs in the near-core regions of stars with a convective core, impacts
directly on the age estimation. Precise aging thus requires a thorough
understanding of the relevant mixing processes for representative
samples of stellar populations. We have not yet reached that stage for massive
core hydrogen burning stars, nor for the core helium burning stars. The
available results on seismically tuned mixing concern too few stars in these
categories so far. However, asteroseismology is well on its way to achieve
appropriate calibrations for mixing processes that are needed to bring the
observational seismic data into agreement with theoretical models.

\section{Interior Rotation}
\label{SectionRotation}

Application of asteroseismology to stars of different mass and rotation is
the best route to make progress in the improvement of stellar physics.  Indeed,
{\it a key ingredient that remained uncalibrated in stellar
  evolution theory until recently is the interior rotation frequency $\Omega(r)$
  during stellar life.}  The angular momentum that stars have at birth, after
the completion of the various stages of their formation process, is essentially
unknown.  The poorly understood star--to--disk interaction and wind and
accretion properties of protostars during the contraction phase imply that we
cannot quantify the rotation profile $\Omega(r)$ at birth from first
principles. Further, the redistribution of angular momentum during the life of
the star is critical for its evolution, yet hard to evaluate in practice.
Asteroseismology is the best way to deliver constraints on $\Omega (r)$.  While
this was already achieved from long-term monitoring of a few unevolved p-mode
pulsators among massive B-type 
stars from ground-based data (Aerts et al.\ 2003; Pamyatnykh et al.\ 2004;
Briquet et al.\ 2007), leading to estimates of the core-to-envelope rotation
$\Omega_{\rm core}/\Omega_{\rm env}$ between 1 and 4, major progress was made
recently on the basis of {\it Kepler\/} light curves.

While rotation affects the period spacings of g-modes as illustrated in
Fig.\,\ref{Timothy}, its clearest signature in any pulsation spectrum is the
occurrence of rotationally split multiplets. The frequency of each mode of
degree $\ell$ gets split into $2\ell+1$ multiplet components due to rotation.
Depending on the inclination angle between the rotation axis of the star and the
line-of-sight, and on the amplitudes with which the individual multiplet
components get excited, all or only a few of the multiplet components can be
detected. To achieve this, the time series must be sufficiently long to have
enough resolving power for the frequency multiplet detections in the case of
slow rotators. This was achieved for 19 g-mode triplets of the SPB star
KIC\,10526294, leading to a rotation period of 188\,d (P\'apics et al.\ 2014)
and hints of the presence of counter-rotation in its envelope (Triana et al.\
2015), from four years of {\it Kepler\/} monitoring.  
Frequency multiplets of low-frequency g-modes easily get merged for intermediate
to fast rotators or they may even lead to power in different frequency regimes
for prograde versus retrograde modes, as is the case for the $\gamma\,$Dor star
in Fig.\,\ref{Timothy} whose $v\sin\,i=72\pm 8\,$km\,s$^{-1}$ (Tkachenko et al.\
2013).
\begin{figure}
\begin{center}
\rotatebox{270}{\resizebox{8.cm}{!}{\includegraphics{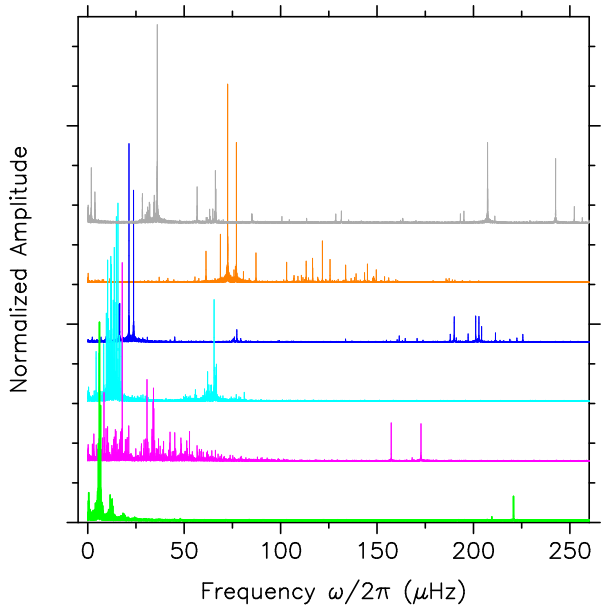}}}
\end{center}
\caption[]{Pulsation spectra of six intermediate-mass F-type hybrid pulsators in
  the core hydrogen burning stage observed with the {\it Kepler\/} satellite.
  From top to bottom: KIC\,4749989, KIC\,1849235, KIC\,8645874, KIC\,9751996,
  KIC\,6468146, KIC\,6367159.  The dimensionless amplitudes are the measured
  amplitudes divided by the amplitude of the strongest mode, shifted along the
  $y$-axis for visibility purpose.  For each of these stars, hundreds of
  significant frequencies were detected in the shown frequency range.}
\label{hybrids}
\end{figure}

A major asset to deduce rotational information for young unevolved intermediate
mass stars are {\it hybrid pulsators}. These are stars
that undergo both p-mode and g-mode pulsations, delivering seismic probing power
throughout the entire star. Figure\,\ref{hybrids} shows the frequency spectra of
six such hybrid F-type pulsators observed with {\it Kepler\/} and exhibiting
pulsational signal in a broad frequency range spanning some 250\,$\mu$Hz.  A few
hybrid pulsators have recently allowed to estimate $\Omega_{\rm core}$ and
$\Omega_{\rm env}$. This research was pioneered by Kurtz et al.\ (2014) and Saio
et al.\ (2015), who discovered rotationally split triplets and quintuplets in
two F-type hybrids and deduced average rotation periods of about 100\, and 65\,d
for them, one star having a slightly faster envelope than core rotation
(KIC\,11145123) while the opposite is the case for the other
(KIC\,9244992). These $\Omega_{\rm core}/\Omega_{\rm env}$ are
essentially model-independent whenever p- and g-mode multiplets are detected.

\begin{figure*}
\begin{center}
\rotatebox{270}{\resizebox{9.8cm}{!}{\includegraphics{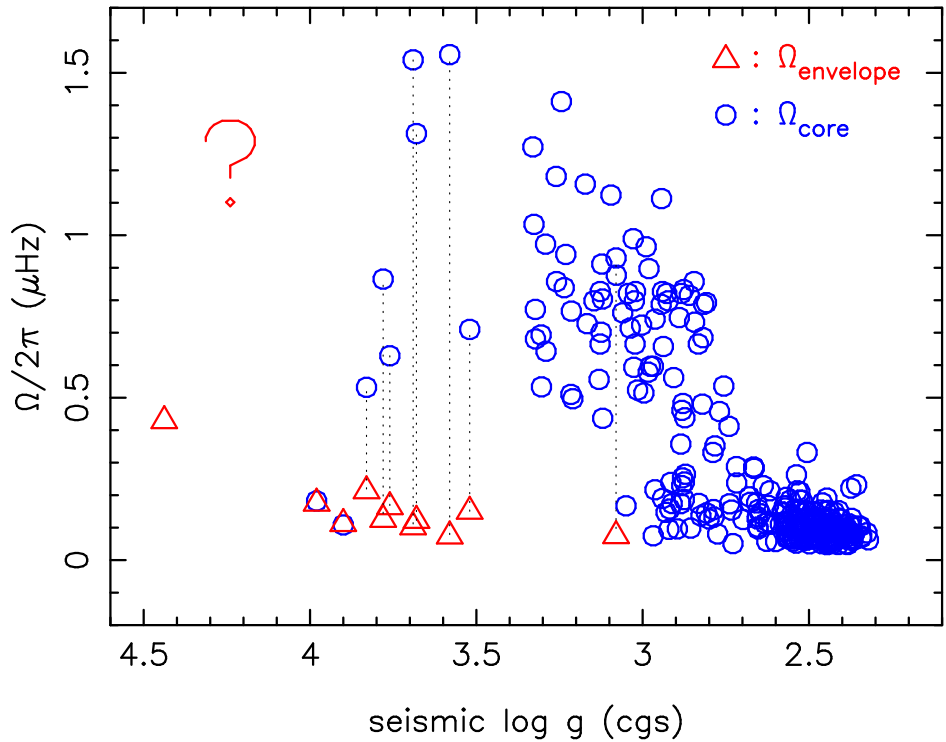}}}
\end{center}
\caption[]{Core rotation frequency of stars with mass between 0.7 and
  3\,M$_\odot$ in terms of their seismically determined surface gravity
  ($\log\,g$) as a rough proxy of their evolutionary stage. The error on the
  frequencies is smaller than the symbol size, while 0.05 is a typical
  uncertainty for the seismic $\log\,g$ values.  The leftmost triangle
  represents the envelope rotation of the Sun, averaged over latitude;
  $\Omega_{\rm core}$ is not available for the Sun.  Stars with both
  $\Omega_{\rm core}$ and $\Omega_{\rm env}$ have these values connected by a
  dotted line.  Figure constructed from data in Mosser et al.\ (2012), Deheuvels
  et al.\ (2012, 2014), Beck et al.\ (2014), Kurtz et al.\ (2014), Saio et al.\
  (2015).}
\label{Core-rotation}
\end{figure*}

The rotationally-split multiplet components of mixed dipole modes have similar
probing power than in the case of hybrids with multiplets.  Their ability to
derive constraints on $\Omega (r)$ was first exploited by Beck et al.\ (2012),
who studied several red giants after two years of {\it Kepler\/} monitoring.  A
large sample of red giants was meanwhile analysed by Mosser et al.\
(2012). These authors showed that dipole mixed modes mainly allow to deduce the
core rotation, although inversion of the rotationally split multiplets in
subgiants and giants in the early red giant branch also delivered limits on the
envelope rotation (Deheuvels et al.\ 2012, 2014; Beck et al.\ 2014).  We have
$\Omega_{\rm core}$-values
available for hundreds of subgiants and red giants determined by
Mosser et al.\ (2012). All stars of masses between 0.7 and 3\,M$_\odot$ whose
interior rotation frequency was measured seismically have been assembled in
Fig.\,\ref{Core-rotation}. These seismic data shed new light on the theory of
stellar evolution, because they deviate from it by two orders of magnitude in
the red giant stage (e.g., Eggenberger, Montalb\'an \& Miglio 2012; Cantiello et
al.\ 2014). This implies that at least one physical ingredient that couples the
stellar core to the envelope and that is active prior to the core helium burning
stage is missing, irrespective if the star was born with a convective or
radiative core.  Earlier on, seismic evidence for the loss of angular momentum
prior to the white-dwarf stage had already been found (Charpinet et al.\
2009). The red giant asteroseismology now shows that  
angular momentum transport must occur prior to the helium core burning stage for
low to intermediate mass stars.  Unfortunately, we do not yet have any insight
into $\Omega_{\rm core}$ at stellar birth, as indicated by the ``?'' in
Fig.\,\ref{Core-rotation}.
\begin{figure*}
\begin{center}
\rotatebox{270}{\resizebox{9.8cm}{!}{\includegraphics{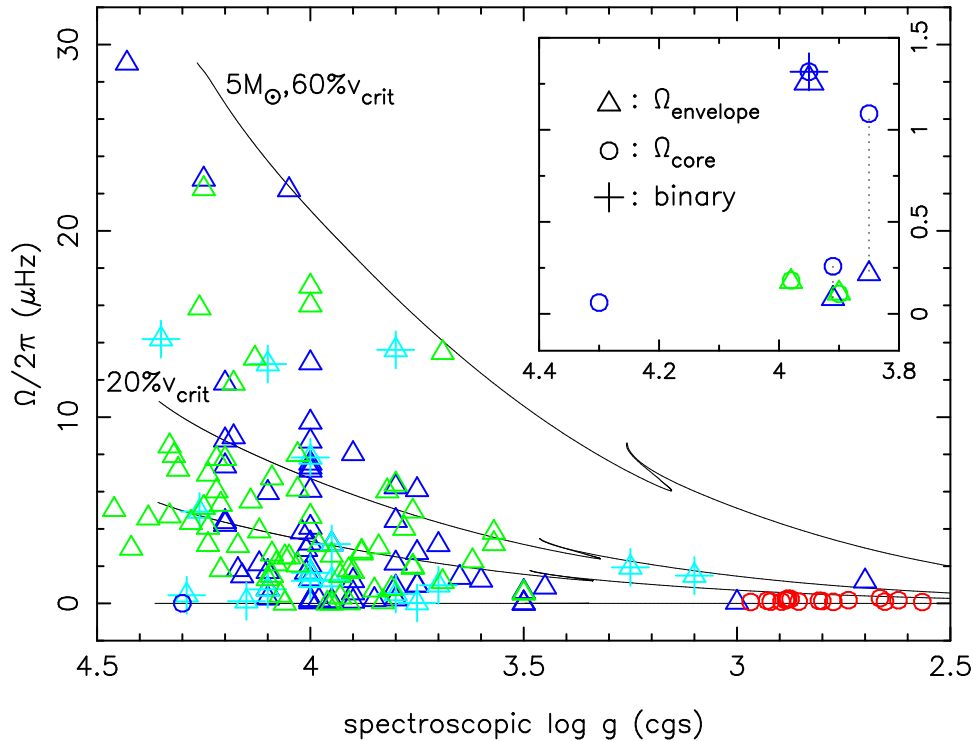}}}
\end{center}
\caption[]{The core and/or envelope rotation frequency of stars with birth mass
  above 2.25 and up to some 50\,M$_\odot$ for which this measurement is
  available, as a function of their spectroscopically determined gravity as a
  proxy for their evolutionary stage. The error on the frequencies is much
  smaller than the symbol sizes. A typical uncertainty for the spectroscopic
  $\log g$ amounts to 0.15. Data for OB stars (dark blue: single; light blue
  with cross: binary) taken from Aerts et al.\ (2014), for Bp/Ap stars from
  Hubrig et al.\ (2007), for Vega from Petit et al.\ (2010), and for the two
  hybrid pulsators from Kurtz et al.\ (2014) and Saio et al.\ (2015)
  (green). The core helium burning red giants with a mass above 2.25\,M$_\odot$
  from Mosser et al.\ (2012) are indicated in red.  The inset shows the massive
  stars with seismically determined values of $\Omega_{\rm core}$.  Four
  evolutionary tracks for a 5\,M$_\odot$ star rotating at the indicated
  fractions of the cricital value (the two lower ones valid for 10\% critical
  and without rotation) were taken from Brott et al.\ (2011a).}
\label{OBA-rotation}
\end{figure*}

To place the seismic results in Fig.\,\ref{Core-rotation} into a more global
evolutionary picture and stress that the sample in Fig.\,\ref{Core-rotation}
concerns only slow rotators and low- to intermediate-mass stars, we show in
Fig.\,\ref{OBA-rotation} several samples of unevolved galactic OBA-type stars
with measured $\Omega_{\rm env}$, either from spectro-polarimetry or from
asteroseismology.  In case both a spectro-polarimetric and a seismic measurement
of $\Omega_{\rm env}$ are available, they are fully compatible with each other.
Figure\,\ref{OBA-rotation} contains all the intermediate and high-mass stars
that also have a seismically determined $\Omega_{\rm core}$.  The red giants
that did not undergo a helium flash at the onset of core helium burning but
started it quietly after crossing the Hertzsprung gap are repeated from
Fig.\,\ref{Core-rotation}; all of these also had a convective core during the
core hydrogen burning just as the other shown stars.  Evolutionary tracks for a
5\,M$_\odot$ star rotating at a few fractions of the critical velocity at birth
and based on angular momentum transport in a diffusive approximation while
including rotationally-induced mixing are shown to guide the eye; these models
were taken from Brott et al.\ (2011a).  Despite the fact that the sample shown
in Fig.\,\ref{OBA-rotation} is limited to stars for which $\Omega_{\rm env}$
could be measured, which excludes the fastest rotators as well as stars of much
lower metallicity than the Sun, it covers a much broader range in surface
rotation frequency than those in Fig.\,\ref{Core-rotation}. There is fairly good
agreement between the measured $\Omega_{\rm env}$ and the model tracks for
various masses and rotation values given in Brott et al.\ (2011a). A general
trend is that most of the massive stars for which $\Omega_{\rm env}$ could be
measured, support the picture that their surface gets slowed down efficiently by
the time they have exhausted their central hydrogen, somewhat independently of
the surface rotation with which they were born.  On the other hand, the
processes that transport angular momentum in stars may also induce mixing and
the occurrence of slowly rotating, nitrogen-enriched stars seems to point to
shortcomings in the models in this respect (Brott et al.\ 2011b; Aerts et al.\
2014).

Just as for red giants, stronger core-to-envelope coupling than foreseen in
current state-of-the-art models, again with two orders of magnitude, is required
to bring the models of massive stars in agreement with measured rotation rates
of young neutron stars and white dwarfs (e.g., Langer 2012). It is thus tempting
to assume that one efficient physical angular momentum transport mechanism is at
work in all stars, yet is missing in current models.  Given that the incidence
of magnetic fields is, at best, limited to 10\% for O to F-type stars (e.g.,
Donati \& Landstreet 2009; Petit et al.\ 2013), an excellent candidate mechanism
could be internal gravity waves (e.g., Talon \& Charbonnel 2005; Rogers et al.\
2013). High-precision CoRoT and {\it Kepler\/} space photometry of the most
massive stars supports this suggestion (e.g., Tkachenko et al.\ 2014 for a
discussion).

\section{Tidal Asteroseismology}

None of the above considered pulsations in
close binaries.  Asteroseismic studies of binaries, in particular eclipsing
ones, are an asset for stellar physics because the binarity offers
model-independent constraints on the masses and radii of the stars, while these
parameters are outcomes of seismic modelling in the case
of single stars.  Isochrone fitting
of double-lined eclipsing binaries offers an independent way to tune interior structure
parameters, such as core overshooting (e.g., Torres 2013).  For young massive
stars this gives compatible estimates with seismic modelling of single pulsators
(e.g., Fig.\ 2 in Aerts 2015). 

Obviously, a combination of binary and asteroseismic modelling has the potential
to put tighter constraints on the stellar structure evaluation if the pulsator
is a member of a binary, even if tides are not involved.  This benefit was
already demonstrated for the $\alpha\,$Cen system (e.g., Miglio \& Montalb\'an
2005) and for the abovementioned sdB pulsator PG\,1336-018 (Van Grootel et al.\
2013) from ground-based asteroseismology and is being applied to space
photometry now as well (e.g., Southworth et al.\ 2011; Frandsen et al.\ 2013,
Maceroni et al.\ 2013, Beck et al.\ 2014; Gaulme et al.\ 2014; Boumier et al.\
2014). A recent overview of asteroseismology in eclipsing binaries with focus on
solar-like pulsations is available in Huber (2014).  It is noteworthy that
binary light curve modelling tools developed in the mmag-precision photometry
era are not able to deal with the current $\mu\,$mag photometric precision. The
space photometry revolution thus triggered the development of new software tools
to handle the binary modelling at the level of the precision of the current data
(Prsa et al.\ 2013; Degroote et al.\ 2013; Bloemen et al.\ 2013).

The subject of {\it tidal asteroseismology}, i.e., stellar modelling based on
tidally excited or tidally affected pulsations, only turned into a practical
science since the availability of the uninterrupted CoRoT and {\it Kepler\/}
lightcurves of either eclipsing binaries with tidally excited g-modes (e.g.,
Maceroni et al.\ 2009; Welsh et al.\ 2011; Hambleton et al.\ 2013; Debosscher et
al.\ 2013; Borkovits et al.\ 2014) or pulsating stars that were initially
thought to be single stars but turned out to be a member of a spectroscopic
binary from follow-up data (e.g., P\'apics et al.\ 2013). In all of those
studies, clear evidence was found that some or all of the g-modes are tidally
triggered, after iterative light curve modelling schemes in
the cases where the binary and pulsational variability are of the same order of
magnitude. Modelling of tidally excitated pulsation modes triggered by dynamic tides
active in eccentric binaries offers the opportunity to put stringent constraints
on the interior structure models of the binary components by exploiting the
tide-generating potential. This was so far only worked out in detail for the
most eccentric pulsating {\it Kepler\/} F-type binary KOI-54 by Fuller \& Lai (2012).

Tidal asteroseismology has immense potential. The photometric observational data
have been assembled and are being analysed in terms of pulsation frequencies and
mode identification while spectroscopic orbital monitoring is ongoing for tens
of systems, by various research teams. Full exploitation of the tide-generating
potential will require new work-intense theoretical developments because the
modelling needs to be tuned to each of the binary pulsators individually.

\section{Future Prospects}

The past five years have delivered tremendous progress in asteroseismology.
Thanks to the long-term uninterrupted high-precision space photometry, we have
moved from the detection of a few tens of pulsation modes in a handful of
solar-like stars to hundreds of modes in thousands of single and binary stars of
very diverse evolutionary stage.  Despite great achievements, some of which
discussed above, the best of asteroseismology for stellar physics is yet to
come. Indeed, few of the four-year {\it Kepler\/} light curves have been
exploited in full detail in terms of forward modelling based on individual
frequency matching. While the used scaling relations for solar-like pulsations
are of great value, they can never capture the fine details of the interior
structure for each and every individual star, as forward seismic modelling
can. The pulsation frequencies are now measured with such high precision that
the forward modelling process necessarily has to follow an iterative scheme,
where improvements on the input physics of the stellar models are mandatory to
bring the data in agreement with the theory. Such scheme can only be automated
to a limited extend and requires detailed tweaking, as it was done so
beautifully in Helioseismology (e.g., Christensen-Dalsgaard 2002). Similar
efforts for stars will allow deep understanding of currently poorly described
physical processes inside stars, several of which inaccessible for the Sun
because it only reveals p-modes and is a relatively unevolved slow rotator.

A badly known piece of information in stellar evolution theory that we have
already briefly touched upon concerns the star formation process, the end of
which is taken as the initial condition for stellar evolution computations.
Star formation is a notoriously difficult subject, both from a theoretical and
an observational point of view. The final phases of the star formation process
not only determine the starting point for the life of the star, but are also
highly relevant for exoplanetary system formation and evolution. While the
contraction of protostars towards hydrogen ignition is known in a global sense
(e.g., Palla \& Stahler 1992), details about the matter accretion and energy
dissipation are yet to be investigated and calibrated. Also here,
asteroseismology has a role to play, given that pre main-sequence (PMS) stars of
intermediate mass are expected to undergo p-mode pulsations (Marconi \& Palla
1998). PMS asteroseismology has hardly been explored so far, because such
targets did not occur in the {\it Kepler\/} nominal field-of-view and they were
poorly sampled with CoRoT. Nevertheless, Zwintz et al.\ (2014) recently came up
with the first effort to age PMS stars seismically, by considering the highest
detected p-mode frequency in MOST (Matthews 2007) and CoRoT data of young open
cluster stars as a good proxy for the {\it acoustic cutoff frequency}. This
theoretical quantity is tightly connected with the way the p-modes are damped by
outwardly propagating running waves in the upper atmosphere of the star (e.g.,
Hansen et al.\ 1985) and may thus tune the physical conditions in the atmosphere
of PMS stars. As a side note, we remark that a similar quantity applies at the
other extreme of the longest detectable g-mode period (Townsend 2000a,b). As
notably visible in Fig.\,\ref{Timothy}, rotation strongly affects this g-mode
period cutoff. This has so far remained unexploited in practice for g-mode
pulsators.  To investigate if the PMS phase is of importance for the life of
stars and their planets, such stars have been proposed for observations with the
K2 mission (Howell et al.\ 2014).

Although K2 can only deliver light curves with a duration of some 75\,d, with
somewhat less precision than the nominal {\it Kepler\/} mission, its great asset
is that it can point to various observing fields in the ecliptic, thus
delivering seismic data for a much larger variety of stars in terms of mass,
metallicity, and evolutionary stage. This holds the potential to cover the
entire Hertzsprung-Russell diagram with asteroseismology based on pulsational
phenomena with mode periodicities ranging from a few minutes to several weeks.
The K2 data cannot deliver $\Omega (r)$ for low-mass stars but offers the
opportunity to determine this quantity for massive stars, from rotational
splitting of their p-modes.

With launch foreseen in 2017, the TESS mission (Ricker et al.\ 2015) will
provide high-precision photometry at 2-minute cadence for stars with $I$
magnitude in the range 4 to 13. TESS will operate during two years and has
all-sky coverage. Its light curves will be limited to typically a month
duration, except for the ecliptic poles, which will be monitored during a full
year. Hence we look forward to even more diversity in asteroseismic
applications, including the Magellanic Clouds!

Even farther in the future, the PLATO mission up for launch in 2024 (Rauer et
al.\ 2014), will deliver long-term high-precision photometry at 50\,s cadence in
white light, but with the capability of two-colour photometry for the brightest
targets. For a modest number of targets, monitoring with fast cadence of seconds
is foreseen. PLATO data will cover two yet to be chosen hugh fields-of-view
(2250 square degrees). Unlike any of the previous space missions, PLATO is
specifically designed to have simultaneous asteroseismology and exoplanet
detection capabilities. PLATO will be a 6-year mission, with two long-duration
pointings of two and three years each, and a final year with step-and-stare
option to (re-)visit various fields all over the sky during several weeks to
months.  Indeed, the best is yet to come.

\acknowledgements

Part of the research included in this manuscript was based on funding from the
Fund for Scientific Research of Flanders (FWO), Belgium under grant agreement
G.0B69.13, from the Research Council of KU\,Leuven under grant GOA/2013/012, and
from the National Science Foundation of the USA under grant No.\,NSF
PHY11-25915.  The author is grateful to the organisers of the 2014 Meeting of
the Astronomische Gesellschaft in Bamberg for the opportunity to present a
plenary talk on asteroseismology at this conference.  She also acknowledges the
staff of the Kavli Institute of Theoretical Physics at the University of
California, Santa Barbara, for the kind hospitality during the 2015 research
programme ``Galactic Archaeology and Precision Stellar Astrophysics'', which
provided a stimulating environment to write the current manuscript.



\begin{thebibliography}{}

\bibitem{} Aerts, C.: 2013, EAS 64, 323

\bibitem{} Aerts, C.: 2015, IAUS 307, 154

\bibitem{} Aerts, C., Christensen-Dalsgaard, J., Kurtz, D.W., 2010, {\it
    Asteroseismology\/}, Astronomy and Astrophysics Library, Springer-Verlag

\bibitem{} Aerts, C., Molenberghs, G., Kenward, M. G., Neiner, C.: 2014, ApJ
  781, 14

\bibitem{} Aerts, C., Thoul, A., Daszynska, J., Scuflaire, R., Waelkens, C.,
  Dupret, M. A., Niemczura, E., Noels, A.: 2003, Science 300, 1926

\bibitem{} Auvergne, M., Bodin, P., Boisnard, L., et al.: 2009, A\&A 506, 411

\bibitem{} Beck, P. G., Bedding, T. R., Mosser, B., et al.: 2011, Science 332, 205

\bibitem{} Beck, P.~G., Hambleton, K., Vos J., et al.: 2014, A\&A 564, AA36

\bibitem{} Beck, P. G., Montalb\'an, J., Kallinger, T., et al.: 2012, Nature 481, 55

\bibitem{} Bedding, T. R., Kjeldsen, Butler, R. P., et al.: 2004, ApJ 549, 105 

\bibitem{} Bedding, T. R., Mosser, B., Huber, D., et al.: 2011, Nature 471, 608

	
\bibitem{} Bedding, T. R., Murphy, S. J., Colman, I. L., Kurtz, D. W.: 2014, in:
  Proc.\ CoRoT Symposium 3 / Kepler KASC-7 joint meeting, arXiv:1411.1883

\bibitem{} Bloemen, S., Degroote, P., Conroy, K., Hambleton, K. M., Giammarco,
  J. M., Pablo, H., Prsa, A.: 2013, EAS 64, 269

\bibitem{} Borkovits, T., Derekas, A., Fuller, J., et al.: 2014, MNRAS 443, 3068

\bibitem{} Bouabid, M.-P., Dupret, M.-A., Salmon, S., Montalb\'an, J., Miglio,
  A., Noels, A.: 2013, MNRAS 429, 2500

\bibitem{} Bouchy, F., Carrier, F.: 2001, A\&A 374, L5

\bibitem{} Boumier, P., Benomar, O., Baudin, F., et al.: 2014, A\&A 564, A34

\bibitem{} Brassard, P., Fontaine, G., Wesemael, F., Hansen, C.~J.:  1992,  
ApJS 80, 369

\bibitem{} Briquet, M., Aerts, C., Baglin, A., et al.: 2011, A\&A 527, 112

\bibitem{} Briquet, M., Morel, T., Thoul, A., Scuflaire, R., Miglio, A., Montalbán, J.,
Dupret, M.-A., Aerts, C.: 2007, MNRAS 381, 1482

\bibitem{} Brott, I., de Mink, S. E., Cantiello, M., et al.: 2011a, A\&A 530,  A115

\bibitem{} Brott, I., Evans, C. J., Hunter, I., et al.: 2011b, A\&A 530, A116

\bibitem{} Cantiello, M., Mankovich, C., Bildsten, L., Christensen-Dalsgaard,
  J., Paxton, B.: 2014, ApJ 788, 93

\bibitem{} Chaplin, W.~J., Basu, S., Huber, D., et al.: 2014, ApJS 210, 1

\bibitem{} Chaplin, W. J., Miglio, A.: 2011, ARA\&A 51, 353

\bibitem{} Charbonnel, C., Talon, S.: 2005, Science 309, 2189

\bibitem{} Charpinet, S., Fontaine, G., Brassard, P.: 2009, Nature 461, 501

\bibitem{} Charpinet, S., Van Grootel, V., Fontaine, G., et al.: 2011, A\&A 530, A3

\bibitem{} Christensen-Dalsgaard, J.: 2002, RvMP 74, 1073

\bibitem{} Christensen-Dalsgaard, J.,  Thompson, M. J.: 2011, IAUS 271, 32

\bibitem{} Corsaro, E., Stello, D., Huber, D., et al.: 2012, ApJ 757 190 

\bibitem{} Debosscher, J., Aerts, C., Tkachenko, A., et al.: 2013, A\&A 556, A56

\bibitem{} Degroote, P., Aerts, C., Baglin, A., et al.: 2010, Nature 464, 259 

\bibitem{} Degroote, P., Conroy, K., Hambleton, K., Bloemen, S., Pablo, H.,
  Giammarco, J., Prsa, A.: 2013, EAS 64, 277

\bibitem{} Deheuvels, S., Garc{\'{\i}}a, R.~A., Chaplin, W.~J., et al.: 2012,
  ApJ 756, 19

\bibitem{} Deheuvels, S., Do{\u g}an, G., Goupil, M.~J., et al.: 2014, A\&A 564, AA27

\bibitem{} Donati, J.-F., Landstreet, J.~D.:  2009, ARA\&A 47, 333

\bibitem{} Dupret, M.-A., Belkacem, K., Samadi, R., et al.: 2009, A\&A 506, 57

\bibitem{} Eggenberger, P., Montalb\'an, J., Miglio, A.: 2012, A\&A 544, L4

\bibitem{} Epstein, C. R., Pinsonneault, M. H.: 2014, ApJ 780, 159

\bibitem{} Frandsen, S., Lehmann, H., Hekker, S., et al.: 2013, A\&A 556, A138

\bibitem{} Fuller, J., Lai, D.: 2012, MNRAS 420, 3126

\bibitem{} Garc{\'{\i}}a, R.~A., Ceillier, T., Salabert D., et al.: 2015, A\&A
  572, AA34

\bibitem{} Gaulme, P., Jackiewicz, J., Appourchaux, T., Mosser, B., et al.:
  2014, ApJ 785, 5

\bibitem{} Gilliland, R. L., Brown, T. M., Christensen-Dalsgaard, J., et al:
  2010, PASP 122, 131

\bibitem{} Hambleton, K. M., Kurtz, D. W., Prsa, A., et al.: 2013, MNRAS 434, 925

\bibitem{} Hansen, C.~J., Winget, D.~E., Kawaler, S.~D.: 1985, ApJ 297, 544

\bibitem{} Hekker, S., Gilliland, R.~L., Elsworth, Y., et al.: 2011,  MNRAS 414,
  2594

\bibitem{} Heber, U.: 2009, ARA\&A 47, 211

\bibitem{} Howell, S.~B., Sobeck, Ch., Haas, M., et al.: 2014, PASP 126, 398

\bibitem{} Huber, D.: 2014, in: Giants of Eclipse: The zeta Aurigae Stars and
  Other Binary Systems, ASSL 408, 169

\bibitem{} Huber, D., Bedding, T.~R., Stello, D., et al.: 2011, ApJ 743, 143

\bibitem{} Hubrig, S., North, P., Sch\"oller, M.: 2007, AN 328, 475

\bibitem{} Kallinger, T., Hekker, S., Mosser, B., et al.: 2012, A\&A 541, AA51

\bibitem{} Kjeldsen, H., Bedding, T. R.: 1995, A\&A 293, 87

\bibitem{} Kjeldsen, H., Bedding, T. R., Butler, R. P,, et al.: 2005, ApJ 635, 1281 

\bibitem{} Koch, D. G., Borucki, .J., Basri, G., et al: 2010, ApJ 713, L79 

\bibitem{} Kurtz, D.~W., Saio, H., Takata, M., Shibahashi, H., Murphy, S.~J.,
  Sekii, T.: 2014, MNRAS 444, 102

\bibitem{} Lebreton, Y., Goupil, M. J.: 2014, A\&A 569, A21

\bibitem{} Maceroni, C., Montalb\'an, J., Gandolfi, D., Pavlovski, K., Rainer,
  M.: 2013, A\&A 552, 60

\bibitem{} Maceroni, C., Montalb\'an, J., Michel, E., et al.: 2009, A\&A 508, 1375

\bibitem{} Marconi, M., Palla, F.: 1998, ApJ 507, L141

\bibitem{} Matthews, J.: 2007, CoAst 150, 333

\bibitem{} Mazumdar, A., Monteiro, M. J. P. F. G., Ballot, J., et al.: 2014, ApJ
  782, 18

\bibitem{} Meibom, S., Barnes, S. A., Platais, I., Gilliland, R. L., Latham,
  D.W.,   Mathieu, R. D.: 2015, Nature 517, 589 

\bibitem{} Metcalfe, T.~S., Creevey, O.~L., Do{\u g}an, G., et al.: 2014, ApJS 214, 27 

\bibitem{} Michel, E., Baglin, A., Weiss, W. W., et al.: 2008, CoAst 156, 73

\bibitem{} Miglio, A., Montalb\'an, J.: 2005, A\&A 441, 615

\bibitem{} Miglio, A., Montalb\'an, J., Noels, A., Eggenberger, P.: 2008, MNRAS
  386, 1487

\bibitem{} Mosser, B., Benomar, O., Belkacem, K., et al.: 2014, A\&A 572, L5

\bibitem{} Mosser, B., Goupil, M.~J., Belkacem, K., et al.: 2012, A\&A 548, AA10

\bibitem{} Palla, F., Stahler, S.~W.: 1993, ARA\&A 418, 414


\bibitem{} Pamyatnykh, A. A., Handler, G., Dziembowski, W. A.: 2004, MNRAS 350, 1022

\bibitem{} P{\'a}pics, P.~I., Moravveji, E., Aerts, C., Tkachenko, A., Triana, S.~A., 
Bloemen, S., Southworth, J.:  2014, A\&A 570, AA8

\bibitem{} P\'apics, P. I., Tkachenko, A., Aerts, C., et al: 2013, A\&A 553, A127

\bibitem{} Petit, P., Ligni\`eres, F., Wade, G. A., et al.: 2010, A\&A 523, 41

\bibitem{} Petit, V., Owocki, S. P., Wade, G. A., et al.: 2013, MNRAS 429, 398

\bibitem{} Prsa, A., Degroote, P., Conroy, K., Bloemen, S., Hambleton, K.,
  Giammarco, J., Pablo, H.: 2013, EAS 64, 259

\bibitem{} Rauer, H., Catala, C., Aerts, C., et al: 2014, ExA 38, 249

\bibitem{} Ricker, G.~R., Winn, J.~N., Vanderspek, R., et al.: 2015, JATIS 1, 014003

\bibitem{} Rogers, T.~ M., Lin, D.~N.~C., McElwaine, J.~N., Lau, H.~H.~B.: 2013,
  ApJ 772, 21

\bibitem{} Saio, H., Kurtz, D.~W., Takata, M., Shibahashi, H., Murphy, S.~J., Sekii, 
T., Bedding, T.~R.:  2015, MNRAS 447, 326

\bibitem{} Southworth, J., Zima, W., Aerts, C., et al.: 2011, A\&A 414, 2413

\bibitem{} Stello, D., Huber, D., Bedding, T.~R., et al.: 2013, ApJ 765, LL41 

\bibitem{} Tkachenko, A., Aerts, C., Yakushechkin, A., et al.: 2013, A\&A 556, 52

\bibitem{} Tkachenko, A., Degroote, P., Aerts, C., et al.: 2014, MNRAS 438, 3093

\bibitem{} Torres, G.: 2013, EAS 64, 87

\bibitem{} Townsend, R.~H.~D., 2000a, MNRAS 318, 1

\bibitem{} Townsend, R.~H.~D., 2000b, MNRAS 319, 289

\bibitem{} Triana, A. S., Moravveji, E., P\'apics, P., Aerts, C., Kawaler,
  S. D., Christensen-Dalsgaard, J.: 2015, ApJ, submitted

\bibitem{} Van Grootel, V., Charpinet, S., Brassard, P., Fontaine, G., Green,
  E. M.: 2013, A\&A 553, 97

\bibitem{} Van Grootel, V., Charpinet, S., Fontaine, G., et al.: 2010b, ApJ 718, L97

\bibitem{} Van Grootel, V., Charpinet, S., Fontaine, G., Green, E. M., Brassard,
  P.: 2010a, A\&A 524, 63

\bibitem{} Van Reeth, T., Tkachenko, A., Aerts, C., et al.: 2015a, A\&A 574, 17

\bibitem{} Van Reeth, T., Tkachenko, A., Aerts, C., et al.: 2015b, ApJS, submitted

\bibitem{} Verma, K., Faria, J.~P., Antia, H. M., et al.: 2014, ApJ 790, 138

\bibitem{} Vu\v{c}kovi\'c, M., Aerts, C., \O stensen, R., Nelemans, G., Hu, Haili,
  Jeffery, C. S., Dhillon, V. S., Marsh, T. R.: 2007, A\&A 471, 605

\bibitem{} Welsh, W. F., Orosz, J. A., Aerts, C,. et al.: 2011, ApJS 197, 4

\bibitem{} White, T. R., Bedding, T. R., Stello, D., Christensen-Dalsgaard, J.,
  Huber, D., Kjeldsen, H.: 2011, ApJ 743, 161

\bibitem{} Zwintz, K., Fossati, L., Ryabchikova, T., et al.: 2014, Science 345, 550 
\end{thebibliography}
\end{document}